\documentclass{article}
\usepackage{fleqn,mprocl,amssymb}
\usepackage{graphicx}

\newcommand{\AmS}{{\protect\the\textfont2
  A\kern-.1667em\lower.5ex\hbox{M}\kern-.125emS}}
\def\be{\begin{equation}}
\def\ee{\end{equation}}
\def\bea{\begin{eqnarray}}
\def\eea{\end{eqnarray}}

\begin{document}

\title{REAL-AXIS SOLUTION OF ELIASHBERG EQUATIONS IN VARIOUS
ORDER-PARAMETER SYMMETRIES AND TUNNELING CONDUCTANCE OF
OPTIMALLY-DOPED HTSC}

\author{Giovanni A. Ummarino, Renato S. Gonnelli and Dario Daghero }

\address{INFM - Dipartimento di Fisica, Politecnico di Torino,\\
c.so Duca degli Abruzzi, 24 - 10129 Torino, Italy}


\maketitle \abstracts{In the present work we calculate the
theoretical tunneling conductance curves of SIN junctions
involving high-$T_{\mathrm{c}}$ superconductors, for different
possible symmetries of the order parameter ($s$, $d$,
$s+\mathrm{i}d$, $s+d$, \emph{anisotropic s} and \emph{extended
s}). To do so, we solve the real-axis Eliashberg equations in the
case of an half-filled infinite band. We show that some of the
peculiar characteristics of HTSC tunneling curves (dip and hump at
$eV>\Delta$, broadening of the gap peak, zero bias and so on) can
be explained in the framework of the Migdal-Eliashberg theory.
The theoretical d$I$/d$V$ curves calculated for the different
symmetries at $T$~=~4~K are then compared to various experimental
tunneling data obtained in optimally-doped BSCCO, TBCO, HBCO, LSCO
and YBCO single crystals. To best fit the experimental data, the
scattering by non-magnetic impurities has to be taken into
account, thus limiting the sensitivity of this procedure in
determining the exact gap symmetry of these materials. Finally,
the effect of the temperature on the theoretical tunneling
conductance is also discussed and the curves obtained at
$T$~=~2~K are compared to those given by the analytical
continuation of the imaginary-axis solution. } \vspace{-10mm}
\section{Introduction}
Both experimental evidences and theoretical arguments suggest
that in high-$T_{\mathrm{c}}$ superconductors (HTSC) the $k$-space
symmetry of the energy gap is different from the isotropic
$s$-wave typical of conventional superconductors
\cite{VanHarlingen}. A $d$-wave gap symmetry with nodes on the
Fermi surface, and some mixed symmetries, have been alternatively
proposed as possible candidates for explaining the sometimes
puzzling features of HTSC.

The experimental determination of the symmetry of the energy gap
is therefore a key step for understanding the nature of the
pairing between electrons. In principle, the most usual and
straightforward approach to this problem should consists in
studying the tunneling conductance of HTSC junctions.

In this paper, we calculate the theoretical tunneling conductance
curves for superconductor~-~insulator~-~normal metal (SIN)
junctions involving an high-$T_{\mathrm{c}}$ superconductor, for
various symmetries of the energy gap. We then compare them to the
experimental curves obtained on several high-$T_{\mathrm{c}}$
materials. For the calculation, we make use of the
Migdal-Eliashberg theory \cite{Eliashberg}. The reasonabless of
this approach is proven, among others, by the fact that, for
suitable choices of the parameters, the calculations well
reproduce several typical features of the experimental tunneling
curves.

\section{The model}
Because of the layered structure of copper-oxide superconductors,
we can restrict ourselves to a single Cu-O layer, thus reducing
the dimensionality of the problem. As a result, the $k$-space
becomes a plane and the Fermi surface is a nearly circular line
on this plane. The quasiparticle wavevectors ${\bf k}$ and ${\bf
k}^{\prime}$ are then completely determined by their modulus
$k_{\mathrm{F}}$ and their azimuthal angles $\phi$ and
$\phi^{\prime }$.

In order to calculate the tunneling conductance curves we solve
the real-axis Eliashberg equations (EE) for the order parameter
$\Delta (\omega ,\phi)$ and the renormalization function $Z(\omega
,\phi)$. The kernels of these coupled integral equations contain
the retarded electron-boson interaction $\alpha ^{2}(\Omega%
,\phi ,\phi ^{\prime })F(\Omega )$ and the Coulomb pseudopotential
$\mu ^{\ast }\left( \phi,\phi^{\prime }\right) $
\cite{JohnRenato,nostro,Rieck}. To allow for various symmetries of
the gap, we suppose that these quantities can be expanded, at the
lowest order, in the following way: \vspace{1mm}
\begin{equation}
\alpha ^{2}(\Omega ,\phi ,\phi ^{\prime })F(\Omega )=\alpha _{{\rm is}%
}^{2}F(\Omega )\psi _{{\rm is}}\left( \phi \right) \psi _{{\rm
is}}\left(
\phi ^{\prime }\right) +\alpha _{{\rm an}}^{2}F(\Omega )\psi _{{\rm an}%
}\left( \phi \right) \psi _{{\rm an}}\left( \phi ^{\prime }\right)
\label{alfa^2}
\end{equation}

\begin{equation}
\mu ^{\ast }(\phi ,\phi ^{\prime })=\mu _{{\rm is}}^{\ast }\psi _{{\rm is}%
}\left( \phi \right) \psi _{{\rm is}}\left( \phi ^{\prime }\right) +\mu _{%
{\rm an}}^{\ast }\psi _{{\rm an}}\left( \phi \right) \psi _{{\rm
an}}\left( \phi ^{\prime }\right) .  \label{mu*}
\end{equation}\vspace{1mm}

Here, the subscripts mean \emph{isotropic} and \emph{anisotropic}
respectively, and $\psi_{is}\left(\phi \right) $ and
$\psi_{an}\left( \phi \right) $ are basis functions defined by :
$\psi _{{\rm is}}\left( \phi \right)=1$;
$\psi _{{\rm an}}\left( \phi \right)=\sqrt{2}\cos \left( 2\phi%
\right)$ for the $d$-wave, $\psi _{{\rm an}}\left( \phi \right)=%
8\sqrt{2/35}\cos ^{4}\left( 2\phi \right)$ for the
\emph{anisotropic s}-wave,  and $\psi _{{\rm an}}\left( \phi \right)=%
-2\sqrt{2/3}\cos ^{2}\left( 2\phi \right)$ for the \emph{extended
s}-wave \cite{VanHarlingen}.

We are interested in solutions of the real-axis EE containing
separate isotropic and anisotropic terms, such as:\vspace{1mm}
\begin{eqnarray}
\Delta (\omega ,\phi ) &=&\Delta _{{\rm is}}(\omega )+\Delta _{{\rm an}%
}(\omega )\psi _{{\rm an}}\left( \phi \right)  \label{DeltaZeta}
\\
Z(\omega ,\phi ) &=&Z_{{\rm is}}(\omega )+Z_{{\rm an}}(\omega )\psi _{{\rm an%
}}\left( \phi \right) .  \nonumber
\end{eqnarray}\vspace{-2mm}

Inserting these expressions for $\Delta(\omega, \phi)$ and
$Z(\omega, \phi)$ in the real-axis EE makes them split into four
equations for $\Delta_{\mathrm{is}}$, $\Delta_{\mathrm{an}}$,
$Z_{\mathrm{is}}$ and $Z_{\mathrm{an}}$. From now on, we will put
$\mu^{*}$=0 for simplicity and we will admit that
$Z_{\mathrm{an}}$ is identically zero, as usually
happens~\cite{JohnRenato}. The three remaining equations will be
reported elsewhere~\cite{nostro}. To further simplify the problem
we will put \mbox{$\alpha _{{\rm an}}^{2}F(\Omega )=g\cdot \alpha _{{\rm%
is}}^{2}F(\Omega )$} where $g$ is a constant~\cite{JohnRenato}.
Then, \mbox{$\lambda _{{\rm is}%
}=2\int_{0}^{+\infty }{\rm d}\Omega \alpha _{{\rm is}}^{2}F(\Omega
)/\Omega $}\hspace{1mm} and \mbox{$\lambda _{{\rm an}}=\left(
1/\pi \right) \int_{0}^{2\pi }{\rm d}\phi \psi _{{\rm
an}}^{2}\left( \phi \right) \int_{0}^{+\infty }{\rm d}\Omega
\alpha _{{\rm an}}^{2}F(\Omega )/\Omega$} result to be
proportional: $\lambda _{{\rm an}}=g\cdot \lambda _{{\rm is}}$.

Once obtained the real-axis solutions $\Delta _{{\rm is}}(\omega )$ and $%
\Delta _{{\rm an}}(\omega )$, we calculate the quasiparticle
density of states: \vspace{3mm}
\begin{equation}
N(\omega )=\frac{1}{2\pi }\int_{0}^{2\pi }d\phi \frac{\omega }{%
\sqrt{\omega ^{2}-\left[ \Delta _{{\rm is}}(\omega )+\Delta _{{\rm an}%
}(\omega )\psi _{{\rm an}}\left( \phi \right) \right] ^{2}}}.
\label{Nis}
\end{equation}
\vspace{3mm}

\noindent{whose convolution with the Fermi distribution function
gives the normalized conductance the main features of which will
be discussed in the following section.}

\section{Theoretical results}
The function $\alpha _{is}^{2}F(\Omega )$, which is needed for
numerically solving the real-axis EE, is actually unknown. As a
first approximation, we use for $\alpha^{2}_{is}F(\Omega)$ the
experimental $\alpha ^{2}F(\Omega )_{{\rm Bi2212}}$ (determined in
a previous paper by starting from tunneling data on Bi2212 break
junctions \cite{PhysicaC97}) suitably scaled to give
$T_{\mathrm{c}}$~=~97~K (which is chosen as a representative value
for HTSC): \vspace{3mm}
\begin{equation}
\alpha _{{\rm is}}^{2}F(\Omega )=\frac{\lambda _{{\rm
is}}}{\lambda }\alpha ^{2}F(\Omega )_{{\rm Bi2212}}.
\label{alfaBi}
\end{equation}
\vspace{3mm}

The details of the shape of $\alpha _{is}^{2}F(\Omega )$ are
actually not relevant for the solutions, which instead strongly
depend on the quantity $\omega _{\log }=$ $\exp \left( \frac{2}{\lambda%
}\int_{0}^{+\infty }d\omega \frac{\alpha ^{2}F\left( \omega%
\right) }{\omega }\ln \omega \right) $.

\begin{figure}[t]
\includegraphics[keepaspectratio,width=12cm]{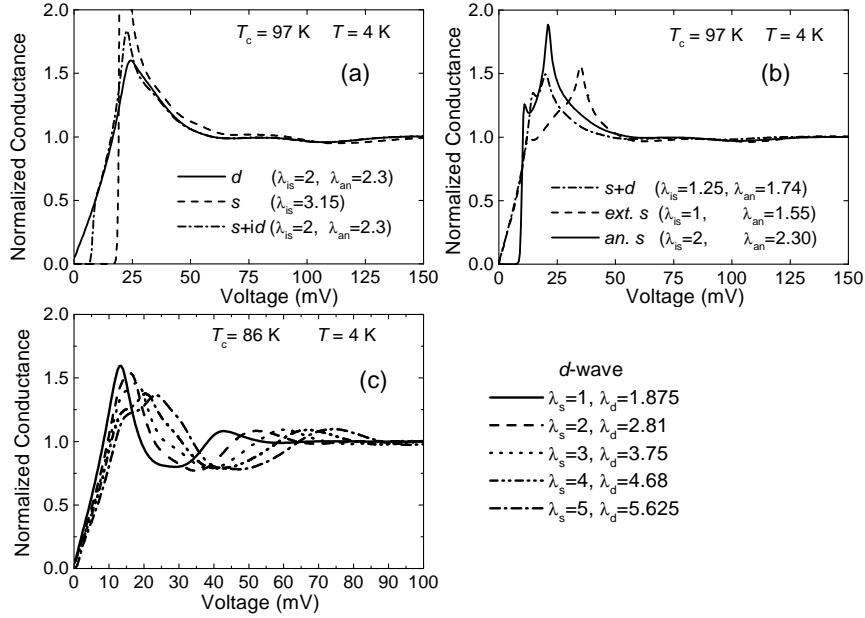}
\caption{(a,b) Theoretical tunneling density of states for
various symmetries at $T$=4 K and $\mu ^{\ast}$=0. The values of
$\lambda_{\mathrm{is}}$ and $\lambda_{\mathrm{an}}$ corresponding
to each curve are indicated. (c) Theoretical normalized tunneling
conductance in the $d$-wave symmetry, for increasing values of the
coupling constants. The dip at about 2$V_{\mathrm{peak}}$ and the
hump at about 3$V_{\mathrm{peak}}$
are well evident.}\label{f:1}
\end{figure}

Figures \ref{f:1}(a) and \ref{f:1}(b) show the theoretical
conductances calculated at the temperature $T=4$~K for all the
symmetries analyzed here: $s$, $d$, $s+\mathrm{i}d$, $s+d$,
\emph{anisotropic s} and \emph{extended s}. For each curve, a
couple of $\lambda_{\mathrm{is}}$ and $\lambda_{\mathrm{an}}$
values is indicated which gives the expected symmetry and the
correct critical temperature. As we shall see in the next section,
these curves well reproduce some of the characteristic features of
HTSC, such as the conductance excess below the gap and the
broadening of the peak. Another apparently anomalous feature of
HTSC recently reported in literature \cite{Ozyuzer} is a
suppression of the conductivity at about 2$V_{\mathrm{peak}}$
(usually known as ``dip'') and a subsequent enhancement at about
3$V_{\mathrm{peak}}$ (``hump''). Both these effects can be easily
obtained by starting from the Eliashberg equations, using high
values of the coupling constants \cite{Varelogiannis} and
reducing $\omega _{\log }$ in order to maintain the critical
temperature close to the experimental value. Figure \ref{f:1}(c)
shows the conductance curves in the $d$-wave symmetry obtained for
increasing values of the coupling constants. Note that the
appearance of the ``hump'' is strictly related to the presence of
the dip, as required by the conservation of the states. For
instance, in the framework of the Eliashberg theory, a similar
effect on the tunneling conductance curves can also be obtained
by simply taking into account the finiteness of the bandwidth
\cite{bandafinita}.

The curves we have discussed so far were calculated at $T$=4~K, in
view of a comparison to the experimental data which are usually
obtained at the temperature of liquid helium. We also calculated
the $N(V)$ curves at lower ($T$=2~K) and higher ($T$=40,~80~K)
temperatures. The curves at $T$=2~K are not much different from
those reported in Figures \ref{f:1}(a) and \ref{f:1}(b), apart
from the fact that some fine structures (such as, for example,
the first peak in the \emph{anisotropic s} curve) are better
distinguishable. Increasing the temperature results, in general,
in a smoothing and broadening of the conductance peak at the gap
edge. Above a temperature of about $T_{\mathrm{c}}/3$ all the
conductances, apart the $s$-wave one, are almost the same.

\begin{figure}[t]
\includegraphics[keepaspectratio,width=13cm]{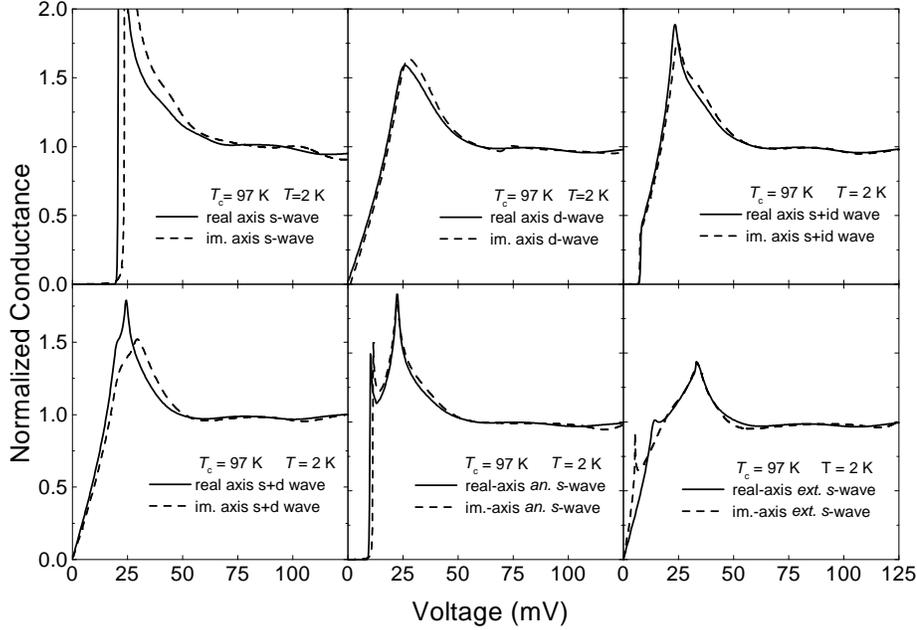}
\caption{Comparison of the normalized tunneling conductance curves
obtained by analytical continuation of the imaginary-axis
solutions of the EE (dashed curves) to those obtained by direct
solution of the real-axis EE (solid curves). All the curves are
calculated at $T$=2~K. The agreement is good for the $d$,
$s+\mathrm{i}d$ and \emph{anisotropic s} symmetries. In the $s$
case, the analytical continuation shifts the gap of about 3 meV
toward higher energies. In the $s+d$ case the peak of the
imaginary-axis solution is lower and broader than that of the
real-axis one. The \emph{extended s} curves agree quite well
except that for very
low voltage.}\label{f:2}
\end{figure}

Just before comparing the theoretical results to the experimental
data obtained on tunnel junctions, let us discuss an important
point concerning the technique here adopted for the solution of
the EE. The most usual and simplest approach to the solution of
these equations consists in facts in solving them in their
imaginary-axis formulation, and then to continue the solutions
$\Delta(\mathrm{i}\omega_{\mathrm{n}})$ and
$Z(\mathrm{i}\omega_{\mathrm{n}})$ to the real axis. Actually, as
we show in Figure \ref{f:2}, the results of this procedure only
\emph{approximately} agree with those obtained by directly solving
the real-axis EE. Note that, in addition, the analytical
continuation is meaningful only at very low temperatures (this is
the reason why the curves in Figure \ref{f:2} are calculated at
$T$=2~K).

\section{Comparison to experiments}
Let us now compare the theoretical conductance curves to those
obtained on optimally-doped single crystals \cite{PhysicaC97,LSCO}
of BSSCO , LSCO, YBCO, TBCO and HBCO . Here we only use for the
comparison the theoretical curves obtained in the $d$ symmetry,
which in fact better agree with experimental data. For each of
these materials we use for $\alpha^{2}_{\mathrm{is}}F(\Omega)$
the phonon spectral density determined by neutron scattering
\cite{G(omega)}, suitably scaled to obtain the best fit. As
anticipated in a previous section, a large zero-bias often
appears in the experimental tunneling curves. To best fit them,
we add in the EE a term taking into account the scattering from
impurities~\cite{nostro}. We then solve the real-axis EE with
varying $\lambda_{is}$ and $\lambda_{an}$, and choose the couple
of values which give the best-fit curve.

\begin{figure}[t]
\includegraphics[keepaspectratio,width=13cm]{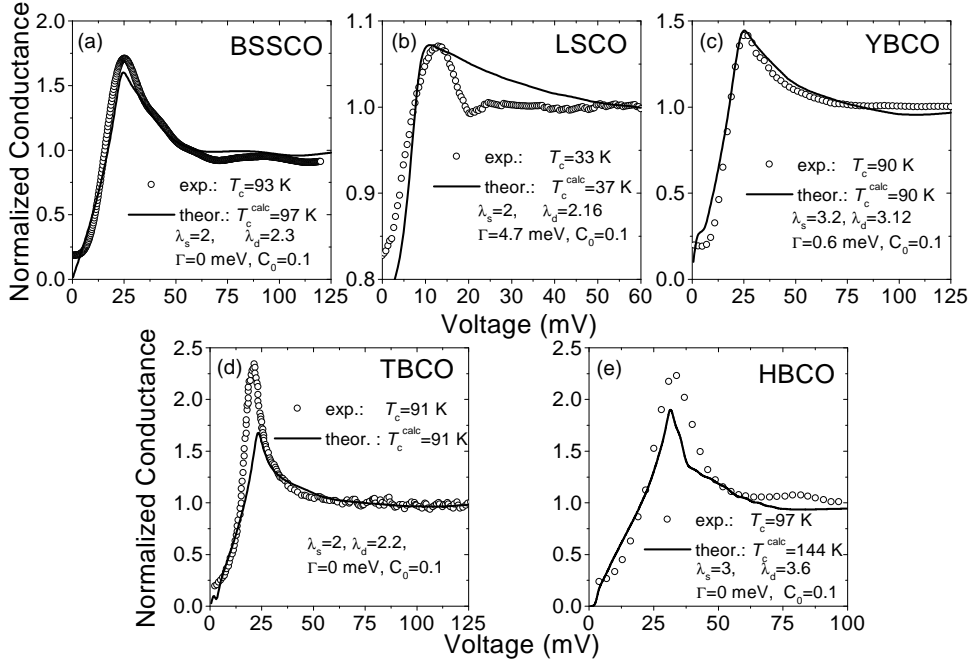}
\caption{Comparison of the theoretical normalized tunneling
conductance curves for the $s+\mathrm{i}d$-wave symmetry to those
experimentally obtained on BSSCO (a), LSCO (b), YBCO (c), TBCO
(d), HBCO (e). In each case the experimental and the calculated
critical temperatures ($T_{\mathrm{c}}$ and
$T_{\mathrm{c}}^{\mathrm{calc}}$) are indicated. Here $\Gamma$ is
proportional to the impurity concentration and
$C_{0}=\cot(\delta_{0})$, where $\delta_{0}$ is the scattering
phase shift.}\label{f:3}
\end{figure}

The results are shown in Figure \ref{f:3}~(a)-(e). In the case of
BSSCO (a) and YBCO (c) the theoretical curves well agree with the
experimental data, and also the calculated critical temperature
is close to the measured one ($T_{\mathrm{c}}$). In the case of
LSCO (b) the observed dip above the gap cannot be reproduced in
the framework of Eliashberg theory. In fact, the only way for
obtaining the great zero-bias value (about 0.8) which appears in
this dataset, is to increase the impurity concentration to such a
large amount that the peak is broadened and the dip cancels out.
For TBCO (d) and HBCO (e), the height of the peak at the gap edge
is quite different in the experimental and theoretical curves.
Actually, this arises from the choice of the normalization for the
experimental data we took from the literature. With this
normalization, in fact, the experimental curves seem to violate
the conservation of the states, which is instead necessarily
obeyed by the theoretical curves. In these cases, the agreement
has to be evaluated by comparing only the \emph{shape} of the
curves, as they were plotted on different scales. Finally, while
for TBCO the experimental and the theoretical critical
temperatures coincide, in the case of HBCO we have
$T_{\mathrm{c}}^{\mathrm{calc}}$=144~K, which is much greater than
the experimental one ($T_{\mathrm{c}}$=97~K)

\section{Conclusions}
We discussed the effect of different possible symmetries of the
order parameter on the tunneling conductance curves of HTSC in the
framework of the Eliashberg theory. We showed that this theory
allows to explain many characteristic features of HTSC which
differ from a classic BCS behaviour, and is therefore a useful
tool to investigate the properties of these materials. We also
tested the theoretical expectations by comparison with
experimental data obtained on many different copper-oxide
superconductors, and we found in many cases a reasonable agreement
between the results of calculation and measurements. The most
probable symmetry of the order parameter seems to be the
$d$-wave, even though the sensitivity of this method in
distinguish between different possible symmetries of the pairing
state is strongly limited by the quality of the crystals.



\end{document}